\documentclass[iop,numberedappendix,appendixfloats]{emulateapj}

\usepackage{amsmath,amssymb,bm,mathrsfs}
\usepackage[symbol]{footmisc}

\newcommand{\apx}[1]{^{\mbox{\tiny{(#1)}}}}

\begin{document}

\def\vk{{\bm{k}}}

\title{A Note on the Radiative and Collisional Branching Ratios\break
in Polarized Radiation Transport with Coherent Scattering}
\author{R.\ Casini,$^a$ T.\ del Pino Alem\'an,$^a$ and R.\ Manso Sainz$^b$}

\renewcommand*{\thefootnote}{\fnsymbol{footnote}}

\affil{$^a$High Altitude Observatory, National Center for Atmospheric
Research,\footnote{The National Center for Atmospheric Research is sponsored
by the National Science Foundation.}
P.O.~Box 3000, Boulder, CO 80307-3000, U.S.A.\break
$^b$Max-Planck-Institut f\"ur Sonnensystemforschung,
Justus-von-Liebig-Weg 3, 37077 G\"ottingen, Germany}

\begin{abstract}
We discuss the implementation of physically meaningful branching ratios
between the CRD and PRD contributions to the emissivity of a polarized 
multi-term atom in the presence of both inelastic and elastic collisions. 
Our derivation is based on a recent theoretical formulation of partially 
coherent scattering, and it relies on a heuristic diagrammatic analysis 
of the various radiative and collisional processes to determine the proper 
form of the branching ratios. 
The expression we obtain for the emissivity is
\begin{displaymath}
\bm{\varepsilon}=\left[\bm{\varepsilon}\apx{1}
		      -\bm{\varepsilon}\apx{2}_{\rm f.s.} \right]
	+\bm{\varepsilon}\apx{2}\;,
\end{displaymath}
where $\bm{\varepsilon}\apx{1}$ and $\bm{\varepsilon}\apx{2}$ are the
emissivity terms for the redistributed and partially coherent radiation,
respectively, and where ``f.s.''\ implies that the corresponding term
must be evaluated assuming a flat-spectrum average of the incident
radiation.
This result is shown to be in agreement with prior literature on the 
subject in the limit of the unpolarized multi-level atom.
\end{abstract}

\section{Introduction}
\label{sec:intro}

\begin{figure}[t!]
\centering
\includegraphics[width=.7\hsize]{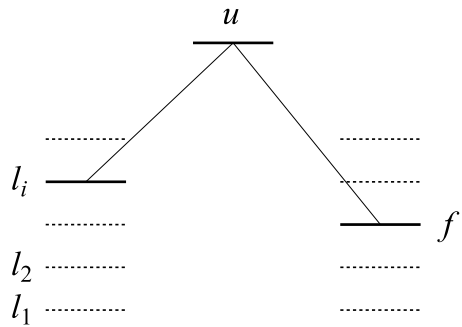}
\caption{Schematic diagram of the $\Lambda$-type multi-term atom.
In order to correctly describe the polarization properties of the outgoing light
in the $u\to f$ transition, all lower terms $l_i$ that are radiatively
or collisionally connected to the upper term $u$, including the final 
term $f$, must be taken into account.
\label{fig:model}}
\end{figure}

The formal theory of spectral line formation in a two-term atom (and 
more generally in a multi-term atom of the $\Lambda$-type; see
Figure~\ref{fig:model})---extended 
perturbatively to fully include second-order atom-photon processes, 
so to be able to describe partial redistribution (PRD) 
effects---predicts that the vector
radiative transfer (RT) equation acquires a new source term, 
$\bm{\varepsilon}\apx{2}$, which describes the \emph{coherent} (in the
broader sense 
of ``memory preserving'') scattering of radiation from the lower 
term \citep{Ca14}. This \emph{second-order emissivity} appears in addition 
to the usual source term $\bm{\varepsilon}\apx{1}$ corresponding to the 
emission of completely redistributed radiation via spontaneous 
de-excitation of the upper state.

In other words, to second order of perturbation, the interaction of 
an atomic system with an incoming beam of photons acquires an 
additional \emph{scattering channel}, beside the one corresponding to the 
pure absorption of a photon with the consequent excitation of the target 
atom (i.e., $\vk+a\to b$, where $\vk$ is the incoming photon, and $a$ 
and $b$ two atomic states satisfying the energy conservation relation 
$\omega_k+\omega_a=\omega_b$). 
%
In the new scattering channel available to perturbative second order, 
the incident photon is instead immediately re-emitted after a \emph{virtual} 
excitation of the atom, possibly leaving the atom in a different 
energy (and polarization) state from the original one
(i.e., $\vk+a\to\vk'+b$, with 
$\omega_k+\omega_a=\omega_{k'}+\omega_b$; the condition
$\omega_a=\omega_b$ evidently corresponds to Rayleigh scattering,
whereas $\omega_a\ne\omega_b$ to Raman scattering).
 
It is verified \citep[cf.][see also \citeauthor{CM16} \citeyear{CM16}]{Ca14} 
that the two emissivity terms $\bm{\varepsilon}\apx{1}$ and
$\bm{\varepsilon}\apx{2}$ evaluate to exactly the same quantity,
in the case of spectrally flat incident radiation, \emph{if one adopts 
the solution of the statistical equilibrium (SE) problem for the 
interacting system restricted to one-photon processes}.
This implies that including both emissivity terms in the second-order
RT while adopting the atomic density matrix solution of the first-order
SE problem would generally lead to double counting the energy radiated by the 
scatterer.

On the other hand, the parallel extension to second-order atom-photon
processes of the SE problem in the formalism of \cite{Ca14} suggests 
that a partial cancellation of the radiative rate for one-photon 
absorption occurs in the presence of coherent scattering, with a 
consequent reduction of the population of 
the excited state with respect to the case of one-photon processes. 
In the limit of \emph{infinitely sharp lower levels}, this population
cancellation is found to be total, in agreement with the simple physical 
argument that no upper state population can be produced when the 
lower state has an infinite lifetime.
One of the expected results of a self-consistent solution of the 
combined SE+RT problem to perturbative second order is that the 
radiative energy is conserved even when both emissivity terms 
$\bm{\varepsilon}\apx{1}$ and $\bm{\varepsilon}\apx{2}$ are taken into account.

Recently, the expression for the RT equation including coherent 
scattering has been applied to modeling various examples of the
partial redistribution of polarized radiation in spectral lines 
from $\Lambda$-type transitions formed in a collisionless plasma
\citep{CM16}. 
Since the formal derivation of the corresponding set of SE equations 
has not yet been completed, the solution of the first-order SE problem 
was adopted in that work.
This creates no formal issues of energy conservation, in the limit of 
infinitely sharp lower 
levels assumed by the model, since no population of the upper state 
from true absorption of the incoming photons is 
expected in that case. Hence, one can assume 
that the scattering of radiation is completely described by 
$\bm{\varepsilon}\apx{2}$, and the contribution of 
$\bm{\varepsilon}\apx{1}$ to 
the RT equation can be omitted altogether. Of course, 
the condition of infinite lifetime of the lower state in deriving the 
first-order SE solution can only be approximated numerically by using 
an extremely diluted radiation field to illuminate the scatterer
\citep{CM16}.

%

\begin{figure}[t!]
\centering
\includegraphics[width=\hsize]{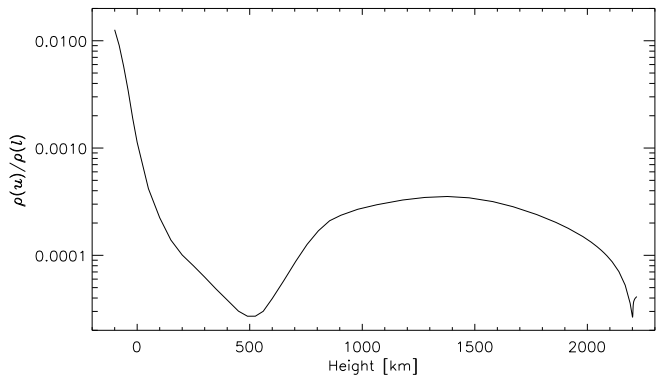}
\caption{\label{fig:rho-trend}
Height dependence of the $\rho(u)/\rho(l)$ ratio for the \ion{Mg}{2} 
two-term model atom in the collisional FAL-C atmosphere.}
\end{figure}

On the other hand, in the application of numerical RT including PRD 
effects to realistic 
models of optically thick atmospheres, the fraction of the upper 
state population $\rho(u)$ that enters the first-order 
emissivity can be significant (typically of the order of 1\% at the 
bottom of the atmosphere; see Figure~\ref{fig:rho-trend}).
This is even more critical in the presence of collisions, 
as it is safe 
to assume that collisional excitation \emph{always} leads to the 
complete redistribution of the energy of the scattered radiation.
Hence, in the absence of a formal self-consistent solution of the 
combined SE+RT problem to perturbative second order, the question 
naturally arises of how to handle \emph{phenomenologically} the 
contribution of spontaneous de-excitation via $\bm{\varepsilon}\apx{1}$ 
alongside with the coherent-scattering emissivity 
$\bm{\varepsilon}\apx{2}$, without impacting energy conservation.

A possible way of dealing with this type of issues is to introduce
appropriate weights between the contributions of $\bm{\varepsilon}\apx{1}$ 
and $\bm{\varepsilon}\apx{2}$ in the RT equation for the Stokes vector 
$\bm{S}\equiv(S_0,S_1,S_2,S_3)^{\rm T}$ of the propagating radiation 
field, i.e.\ \citep[cf.][]{Ca14},
\begin{eqnarray} \label{eq:RT}
\frac{d}{ds}\,\bm{S}(\omega_{k'},\bm{\hat k}') 
&=& -{\bf K}(\omega_{k'},\bm{\hat k}')\,
	\bm{S}(\omega_{k'},\bm{\hat k}') \nonumber \\
&&{}
+\alpha\,\bm{\varepsilon}\apx{1}(\omega_{k'},\bm{\hat k}')
+\beta\,\bm{\varepsilon}\apx{2}(\omega_{k'},\bm{\hat k}')\;,
\end{eqnarray}
where ${\bf K}(\omega_{k'},\bm{\hat k}')$ is the $4\times4$ absorption 
matrix for the outgoing photon of frequency $\omega_{k'}$ and 
propagation direction $\bm{\hat k}'$, $s$ is the linear coordinate 
along the propagation path, 
and finally $\alpha$ and $\beta$ are the (real) weights for the fully 
incoherent (i.e., completely redistributed) and partially coherent 
scattering contributions, respectively.
Because these weights take the form of probability ratios for the 
various processes that determine the excitation state of the atom, 
they are commonly called \emph{branching ratios}.
The choice of the $\alpha$ and $\beta$ weights will generally depend on 
the properties of the physical system at hand. In particular, they will 
take different forms depending on whether the system includes or not 
collisions.

In this work we do not address the problem of establishing physically
consistent branching ratios for the purely radiative case, and simply 
adopt $\alpha=0$ and $\beta=1$ in such case \cite[cf.][]{CM16}. This 
choice is supported by the underlying assumption of weak incident 
radiation, which in turn is consistent with the hypothesis of infinite 
radiative lifetime for the lower state, and with the approximation of 
neglecting stimulated effects.
We must note that this choice does not create any issues when the ratio 
$\rho(u)/\rho(l)$ is important, since in practical cases this always
happens when the system is close to local thermodynamic equilibrium
(LTE), e.g., at the bottom of the atmosphere (see Figure~\ref{fig:rho-trend}).
In such case, the incident radiation field can be assumed to be spectrally 
flat over a very large interval of the atomic transition's frequency,
and so $\bm{\varepsilon}\apx{2}\equiv\bm{\varepsilon}\apx{1}$,
when the first-order density matrix solution is employed. Thus our choice
of branching ratios for the purely radiative case correctly
reproduces the expected CRD regime of the scattered radiation
at LTE.

\begin{figure}[t!]
\centering
\includegraphics[width=\hsize]{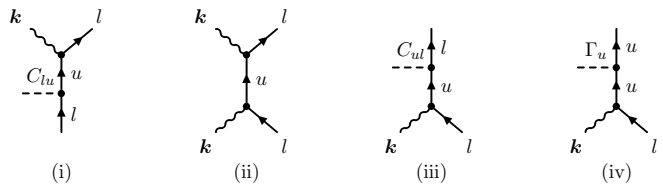}
\caption{\label{fig:diagrams}
Feynman-diagram representation of the dominant second-order 
processes concurrent in a two-level atom in the simultaneous 
presence of radiative and collisional processes (time flows 
upward).}
\end{figure}

We develop our approach to the derivation of branching ratios 
between the CRD and PRD contributions to the RT
equation (\ref{eq:RT}) for the polarized multi-term atom starting 
first from the case of an unpolarized two-level atom. This is a natural 
choice that allows us to directly interpret the branching ratios as event 
probabilities for the various radiative and collisional processes 
that can be realized within the interacting system. 
We use a simple diagrammatic representation of the interacting system
in order to describe the interplay of radiative and 
collisional (both inelastic and elastic) processes. 
After such detailed albeit heuristic analysis, at the end we are able 
to justify very naturally a straightforward extension of our results 
to the case of the polarized multi-term atom.

\section{The case of the unpolarized two-level atom}
\label{sec:coll}

We consider the case of a two-level atom with lower level $l$ and upper
level $u$, in the presence of both radiative and collisional processes. 
%
We have \cite[see, e.g.,][]{LL04}
\begin{eqnarray} \label{eq:rho_coll}
\rho(u)
&\approx& \rho(l)\, 
	\frac{B_{lu} J(\omega_{ul})+C_{lu}}%
	{A_{ul}+B_{ul} J(\omega_{ul})+C_{ul}} \nonumber \\
&=& \rho(u)_{\rm rad} +
	  \rho(u)_{\rm coll}\;.
\end{eqnarray}
where $\rho(u)$ and $\rho(l)$ are the 
relative populations of the upper and lower states, respectively, 
which are subject to the normalization of the trace of the atomic 
density matrix, so that $\rho(u)+\rho(l)=1$.
$J(\omega_{ul})$ is some appropriate average of the ambient 
radiation intensity at the frequency of the atomic transition (see, 
e.g., equation~(\ref{eq:Jdefinition})),
whereas $A_{ab}$ and $B_{ab}$ are the usual Einstein radiative 
coefficients for the transition $a\to b$, and $C_{ab}$ the 
corresponding collisional rate.
We note that the approximate equality in equation~(\ref{eq:rho_coll}) 
becomes exact in the case of an unpolarized 
lower state (see \citealt{LL04}).

Equation~(\ref{eq:rho_coll}) suggests that a fraction
\begin{equation} \label{eq:w_c}
w_c = \frac{\rho(u)_{\rm coll}}{\rho(u)}
	\approx \frac{C_{lu}}{B_{lu} J(\omega_{ul})+C_{lu}}
\end{equation}
of the upper state population is contributed by collisional excitation. 
In that case, there is also a purely collisional contribution to the 
emitted radiation given by $w_c\,\bm{\varepsilon}\apx{1}$. Such 
process is represented schematically by the first diagram of
Figure~\ref{fig:diagrams}.
%
%
In the same figure, diagram (ii) represents the coherent scattering of 
radiation, which is accounted for by the second-order emissivity 
$\bm{\varepsilon}\apx{2}$, and does not contribute to the population of the 
upper state, which is only virtually excited. However, in the presence 
of collisions, the coherence of the scattering process to perturbative 
second order may be destroyed by the intervention of a colliding particle 
(diagrams (iii) and (iv)). This mechanism may either trigger the immediate 
collisional de-excitation of the upper state in the case of an inelastic 
encounter with rate $C_{ul}$ (final state $l$; diagram (iii)), or simply 
redistribute the internal energy of the upper state (with 
\emph{radiatively} induced population 
$\rho(u)_{\rm rad}=(1-w_c)\,\rho(u)$; see equations (\ref{eq:rho_coll})
and (\ref{eq:w_c})) 
in the case of an elastic encounter with rate $\Gamma_u$ (final state $u$; 
diagram (iv)). In the first case, there is no radiative contribution to
equation~(\ref{eq:RT}), whereas in the second case the upper state 
eventually decays with the emission of completely redistributed radiation 
(i.e., associated again with $\bm{\varepsilon}\apx{1}$).

The above analysis should be generalized to properly take into account the 
effects of collisions on atomic polarization. As this
requires the introduction of individual multipolar collisional rates 
$C_{ab}\apx{K}$ in the SE problem, one could think of introducing 
accordingly specific weights $w_c\apx{K}$ for the individual 
multipolar density matrix components $\rho^K_Q(u)$ in 
$\bm{\varepsilon}\apx{1}$. On the other hand, this is quite inconvenient
(as well as ambiguous), 
and an alternative approach will be presented at the very end, after we 
discuss the results of our heuristic diagrammatic analysis in the light 
of other treatments of this problem found in the literature. For the current
discussion, we are assuming that the partition coefficient $w_c$ 
applies identically to all components of the atomic density
matrix, and therefore it is fully appropriate only in the case of the
unpolarized atom.

It is important to observe that when the inelastic collision lifetime of the 
upper state is included in the computation of the damping parameter 
of the line profile, then $\bm{\varepsilon}\apx{2}$ already contains
the proper branching ratio to account for the possibility of collisional 
de-excitation following the radiative excitation of the upper state
(Figure~\ref{fig:diagrams}.(iii)). In order to 
see this, one must recall that the formal derivation of the second-order
emissivity \citep{Ca14} leads to the following expression for the 
atomic-frame redistribution function ${\cal R}(\omega_k,\omega_{k'})$ 
in terms of the generalized line profiles 
$\Psi_{ab,cde}^{\pm k,\pm k'\pm k}$ that appear in 
$\bm{\varepsilon}\apx{2}$ \citep[cf.][equation~(6)]{Ca14},
\begin{eqnarray}	\label{eq:red.def}
&&{\cal R}(\Omega_u,\Omega_{u'};
	\Omega_l,\Omega_{l'},\Omega_f;
	\omega_k,\omega_{k'}) \nonumber \\
&&\kern .2in {}= {\rm i}(\Omega_u-\Omega_{u'}^\ast)\,
(\Psi_{u'l',ful}^{-k,+k'-k} + \bar\Psi_{ul,fu'l'}^{-k,+k'-k})\;,
\end{eqnarray}
where we indicated with $\Omega_a=\omega_a-{\rm i}\epsilon_a$ the 
(complex) frequency of the atomic level $a$, which is comprised of the
true energy $\omega_a$ of the level and its statistical width 
$\epsilon_a$.
Generally, $\epsilon_a$ represents the inverse of the \emph{total} 
lifetime of the quantum state $a$, from which the damping parameter for 
the corresponding energy level is calculated. Therefore, in the presence 
of a non-zero probability for collisional de-excitation, the 
quantity ${\rm i}(\Omega_u-\Omega_{u'}^\ast)$, which appears at the
denominator in the expression of $\bm{\varepsilon}\apx{2}$, is augmented by 
the corresponding collisional rate. Hence, the probability for a 
coherent scattering event to occur is accordingly reduced, and it is
immediately verified that the reduction factor is 
$(1-w_c)=\rho(u)_{\rm rad}/\rho(u)$.

We now recall that $\bm{\varepsilon}\apx{2}\equiv\bm{\varepsilon}\apx{1}$ 
in the absence of collisions and in the limit of spectrally flat 
illumination \citep[cf.][equations~(48) and (49)]{Ca14}, showing 
that $\bm{\varepsilon}\apx{2}$ in that case contains exactly 
$\rho(u)_{\rm rad}\equiv\rho(u)$. Because in the presence of collisions 
the denominator in $\bm{\varepsilon}\apx{2}$ is augmented by $C_{ul}$, 
from equations~(\ref{eq:rho_coll}) and (\ref{eq:w_c}) and the previous 
discussion we conclude that, also in this case, in the limit of 
spectrally flat illumination, $\bm{\varepsilon}\apx{2}$ contains exactly
$\rho(u)_{\rm rad}\equiv(1-w_c)\,\rho(u)$. This demonstrates
that the possibility that the virtually excited upper state
might decay via collisional de-excitation, reducing the contribution 
of coherent scattering to the RT equation (\ref{eq:RT}), is already 
accounted for in the expression of $\bm{\varepsilon}\apx{2}$, and 
that the reduction factor is exactly given by 
$\rho(u)_{\rm rad}/\rho(u)=(1-w_c)$.

Hence, in the presence of radiative and \emph{inelastic} collisional 
processes, $\bm{\varepsilon}\apx{2}$ already provides the correct
contribution of coherent scattering to the emitted radiation
without the need for an additional multiplicative weighting factor. 
The RT equation (\ref{eq:RT}) thus becomes ($\alpha=w_c$, 
$\beta=1$),
\begin{eqnarray} \label{eq:RT.in-coll}
\frac{d}{ds}\,\bm{S}(\omega_{k'},\bm{\hat k}') 
&=& -{\bf K}(\omega_{k'},\bm{\hat k}')\,
	\bm{S}(\omega_{k'},\bm{\hat k}') 
	\nonumber \\
&&{}
+w_c\,\bm{\varepsilon}\apx{1}(\omega_{k'},\bm{\hat k}') 
	+\bm{\varepsilon}\apx{2}(\omega_{k'},\bm{\hat k}')\;,
\end{eqnarray}
where we also recalled equation~(\ref{eq:w_c}).
We note that, in the limit of spectrally flat illumination, 
the above equation properly converges to the equation
for the polarized RT in the CRD regime, since 
$\bm{\varepsilon}\apx{2}=(1-w_c)\,\bm{\varepsilon}\apx{1}$ in that case.

In the additional presence of \emph{elastic} collisions, the same analysis 
as before applies, but now a fully redistributed radiation component 
is contributed also by $\rho(u)_{\rm rad}$, according to our diagrammatic 
analysis presented above (Figure~\ref{fig:diagrams}.(iv)).
Since the main effect of elastic collisions on the line shape comes from 
the perturbation of the energy of the atomic levels, which statistically 
can be approximated with a Lorentzian distribution, it is sensible to
add the elastic collision rate to the damping parameters in the
redistribution function.
%

Similarly, the effect of an elastic collision on an atomic level can be 
interpreted as a process where the atomic state before the collision is
destroyed and immediately recreated into a new (iso-energetic) perturbed 
state, with a characteristic inverse lifetime $\frac{1}{2}\Gamma_u$ for 
the process. 
Accordingly, the energy of the atomic level acquires an additional 
imaginary contribution due to elastic collisions, i.e.,
$\Omega_a=\omega_a-{\rm i}(\epsilon_a+\frac{1}{2}\Gamma_a)$,
and the probability for the coherent scattering of the incident
radiation is once again reduced by the corresponding modification 
of the denominator ${\rm i}(\Omega_u-\Omega_{u'}^\ast)$ in the 
expression of $\bm{\varepsilon}\apx{2}$.
Because the inclusion of elastic collisions to the first-order
SE problem does not affect the population balance between the upper and 
lower states, the proposed modification of $\bm{\varepsilon}\apx{2}$, in the 
limit of spectrally flat illumination, is formally equivalent to the 
substitution ($\Omega_u=\Omega_{u'}$, for the population)
\begin{equation} \label{eq:subst}
\rho(u)_{\rm rad}\;\to\;(1-\gamma)\,\rho(u)_{\rm rad}
\end{equation}
where
\begin{eqnarray} \label{eq:gamma}
\gamma&=& \frac{\Gamma_u}{A_{ul}+B_{ul}J(\omega_{ul})+C_{ul}+\Gamma_u}\;.
\end{eqnarray}
is the probability that the excited atom undergoes an elastic 
encounter with a colliding particle.
This is exactly what we expect from our diagrammatic analysis (see
Figure~\ref{fig:diagrams}.(iv)).
Therefore, even in the additional presence of elastic collisions, 
when the corresponding rate is added to the lifetime $\epsilon_u$ 
of the upper level, $\bm{\varepsilon}\apx{2}$ describes the 
proper contribution of coherent scattering to the emitted radiation 
without the need of an ad-hoc multiplicative branching ratio. 

In turn, according to our diagrammatic analysis, the probability 
(\ref{eq:gamma}) that 
an elastic collision occurs after the atom has been radiatively excited
must bring a new contribution of fully redistributed radiation to the RT
equation (\ref{eq:RT.in-coll}), which is proportional to
\begin{displaymath}
\gamma\rho(u)_{\rm rad}=\gamma(1-w_c)\,\rho(u)\;.
\end{displaymath}

Therefore, \emph{in the general presence of radiative and collisional
processes, both inelastic and elastic}, the RT equation for partially 
coherent scattering by an unpolarized two-level atom becomes,
\begin{eqnarray} \label{eq:RT.coll}
&&\frac{d}{ds}\,\bm{S}(\omega_{k'},\bm{\hat k}') 
=-{\bf K}(\omega_{k'},\bm{\hat k}')\,
	\bm{S}(\omega_{k'},\bm{\hat k}') \nonumber \\
&&\kern .1in{}+ \bigl[w_c + \gamma(1-w_c) \bigr]\,
		\bm{\varepsilon}\apx{1}(\omega_{k'},\bm{\hat k}')
	+\bm{\varepsilon}\apx{2}(\omega_{k'},\bm{\hat k}')\;.
\end{eqnarray}
%
%
%



\section{The case of the unpolarized multi-level atom of the $\Lambda$-type}
\label{sec:multiterm}

So far we have concerned ourselves with establishing the choice of
branching ratios that allows us to apply the formalism of \cite{Ca14} to
the polarized line formation in two-level atoms. \cite{CM16} have shown
that such formalism naturally extends to the treatment of the general
multi-level atom of the $\Lambda$-type (see Figure~\ref{fig:model}),
and so we now want to extend the development of the previous section 
to such more general model atom. In the spirit of the former
development, we do this by considering first the
extension to the unpolarized multi-level atom of the $\Lambda$-type.

We therefore consider an atom with an upper level $u$ connected to a 
set of lower levels ${l_1,l_2,\ldots,l_n}$, which are assumed to be 
isolated from each other, both radiatively and collisionally (see
Figure~\ref{fig:model}).
Then, the first-order solution for the population of the upper state 
is given by (cf.\ equation~(\ref{eq:rho_coll}))
\begin{equation} \label{eq:rhou_one}
\rho(u)\approx\rho(l_i)\,\frac{B_{l_iu}J(\omega_{ul_i})+C_{l_iu}}%
	{A_{ul_i}+B_{ul_i}J(\omega_{ul_i})+C_{ul_i}}\;,\quad\forall
i=1,\ldots,n\;.
\end{equation}
(Once again, we note that the above relation is exact in the case 
of an atom with unpolarized lower states.)
We can therefore write
\begin{eqnarray} \label{eq:rhou_mt}
\rho(u)
&\approx&\sum_{i=1}^n \alpha_i\,
	\rho(l_i)\,\frac{B_{l_iu}J(\omega_{ul_i})+C_{l_iu}}%
	{A_{ul_i}+B_{ul_i}J(\omega_{ul_i})+C_{ul_i}} \nonumber \\
&=&\sum_{i=1}^n \alpha_i\,
	\rho(l_i)\,\frac{B_{l_iu}J(\omega_{ul_i})}%
	{A_{ul_i}+B_{ul_i}J(\omega_{ul_i})+C_{ul_i}} \nonumber \\
&&\kern .2in{}+\sum_{i=1}^n \alpha_i\,
	\rho(l_i)\,\frac{C_{l_iu}}%
	{A_{ul_i}+B_{ul_i}J(\omega_{ul_i})+C_{ul_i}} \nonumber \\
&=& \rho(u)_{\rm rad} +
	  \rho(u)_{\rm coll}\;,
\end{eqnarray}
for any choice of the weights $\alpha_i$ such that
$\sum_i \alpha_i = 1$.

The above equation demonstrates that it is still possible to operate 
a separation between the radiative and collisional contributions to 
$\rho(u)$ even for the more general model of the $\Lambda$-type 
multi-level atom.
This allows us to define $w_c=\rho(u)_{\rm coll}/\rho(u)$ like before 
(cf.\ equation~(\ref{eq:w_c})), and to extend the use of 
equation~(\ref{eq:RT.coll}) to such a model, after performing the 
formal substitutions
\begin{eqnarray} \label{eq:subst2}
&&A_{ul}\to\sum_i A_{ul_i}\;,\quad
B_{ul}J(\omega_{ul})\to\sum_i B_{ul_i}J(\omega_{ul_i})\;, \nonumber \\
&&C_{ul}\to\sum_i C_{ul_i}\;,
\end{eqnarray}
in equation~(\ref{eq:gamma}).
Using equations~(\ref{eq:rhou_one}) and (\ref{eq:rhou_mt}), we thus
arrive at the following generalization of equation~(\ref{eq:w_c}),
\begin{equation} \label{eq:w_c.mt}
w_c=\sum_{i=1}^n
\alpha_i\,\frac{C_{l_iu}}{B_{l_iu}J(\omega_{ul_i})+C_{l_iu}}\;.
\end{equation}

If we limit ourselves to considering the first-order SE+RT problem, 
the weights $\alpha_i$ can be \emph{arbitrarily} chosen, in so far that 
they satisfy the normalization condition $\sum_i\alpha_i=1$.
In contrast, in the application of the branching-ratio formalism 
to the treatment of partially coherent scattering, the choice of these
weights is subject to additional physical constraints, because the
second-order emissivity already accounts for the proper branching 
among all possible transitions $(u,l_i)$.

Thus, in order to determine the proper expression for the weights $\alpha_i$,
we simply observe that the choice
\begin{equation} \label{eq:alpha_i}
\alpha_i=\frac{A_{ul_i}+B_{ul_i}J(\omega_{ul_i})+C_{ul_i}}%
	{\sum_n \left[A_{ul_n}+B_{ul_n}J(\omega_{ul_n})+C_{ul_n}\right]}
\end{equation}
%
%
ensures that the expression for $\rho(u)_{\rm rad}$ in
equation~(\ref{eq:rhou_mt}) becomes formally equivalent to the 
one that is explicitly contained in 
$\bm{\varepsilon}\apx{2}$, when the multi-level atom is
illuminated by a spectrally flat radiation
(see the discussion leading to equation~(\ref{eq:RT.in-coll})). 
We therefore propose equation~(\ref{eq:alpha_i}) as the proper definition
of the weights $\alpha_i$ to be used in the generalized expression
(\ref{eq:w_c.mt}) for $w_c$.

\section{Comparison with previous results}

The above results were completely derived within the framework of a 
recent theory of partially coherent scattering from multi-term atoms in a
collisionless plasma developed by \citeauthor{Ca14} (\citeyear{Ca14}; 
see also \citealt{CM16}). 
%
It is therefore important to show how those results compare with existing 
work on the modeling of the partial redistribution of \emph{unpolarized} 
radiation in multi-level atoms \emph{and} in the presence of collisions.

For our comparison, we rely on the work of \cite{Ui01}, which has become
a standard reference for the numerical modeling of multi-level systems
including PRD effects. We report here only the essential formulas
from that work, adapted to our notation.

We consider a multi-level atom of the $\Lambda$-type, 
consisting of one upper state $u$ and a set
$l_1,l_2,\ldots,l_n$ of lower states (cf.\ Section~\ref{sec:multiterm}
and Figure~\ref{fig:model}).
The intensity emissivity in the $(u,f\equiv l_i)$ branch of the multi-level 
system of transitions can be written as \citep[][equation~(2)]{Ui01}
\begin{equation} \label{eq:RHepsilon}
\epsilon_{i}(\omega_{k'},\bm{\hat k}') = \frac{\hbar\omega_{k'}}{4\pi}\,
\mathcal{N}\rho(u)A_{ul_i}\,
    \psi_i(\omega_{k'},\bm{\hat k}')\;,
\end{equation}
where $\cal N$ is the atomic density, and the generalized line profile 
$\psi_i$ is given by \citep[][equation~(6)]{Ui01}
\begin{widetext}
\begin{eqnarray}\label{eq:RHpsi}
\psi_i(\omega_{k'},\bm{\hat k}')
=\phi_i(\omega_{k'},\bm{\hat k}') 
&+& 
    \frac{1}{\rho(u)}\,
    \frac{\sum_j\rho(l_j)B_{l_ju}}%
	{\sum_n\left[A_{ul_n} 
    + B_{ul_n}J(\omega_{ul_n})+ C_{ul_n}\right] 
	+ \Gamma_u} \nonumber \\
&\times& \oint
    \frac{d\bm{\hat{k}}}{4\pi}\int d\omega_k
\left[\mathcal{R}(\omega_k,\bm{\hat{k}};\omega_{k'},\bm{\hat k}') 
- \phi_j(\omega_k,\bm{\hat{k}})\,
  \phi_i(\omega_{k'},\bm{\hat k}')\right] I(\omega_k,\bm{\hat{k}})\;.
	\kern .5cm
\end{eqnarray}
\end{widetext}
%
%
After substitution of equation~(\ref{eq:RHpsi}) into 
equation~(\ref{eq:RHepsilon}), and making use of the definition
\begin{equation} \label{eq:Jdefinition}
J(\omega_{ul_j}) = \oint\frac{d\bm{\hat{k}}}{4\pi}
	\int d\omega_k\,
	\phi_j(\omega_k,\bm{\hat{k}})\,I(\omega_k,\bm{\hat{k}})\;,
\end{equation}
for the mean intensity of the incident radiation, we find
\begin{widetext}
\begin{eqnarray}\label{eq:RHPRDepsilon}
\epsilon_{i}(\omega_{k'},\bm{\hat k}')
&=& \frac{\hbar\omega_{k'}}{4\pi}\,\mathcal{N}\rho(u)
    A_{ul_i}\,\phi_i(\omega_{k'},\bm{\hat k}') 
	\left\{1 - \frac{1}{\rho(u)}\,\frac{\sum_j\rho(l_j)
    B_{l_ju}J(\omega_{ul_j})}{\sum_n\left[A_{ul_n} 
     + B_{ul_n}J(\omega_{ul_n})+ C_{ul_n} \right] + \Gamma_u}\right\}
\nonumber \\
&+& \frac{\hbar\omega_{k'}}{4\pi}\,
    \frac{\mathcal{N}A_{ul_i}\,
	\sum_j\rho(l_j) B_{l_ju}}{\sum_n\left[A_{ul_n} 
     + B_{ul_n}J(\omega_{ul_n})+ C_{ul_n} \right] + \Gamma_u} 
\oint\frac{d\bm{\hat{k}}}{4\pi}\int d\omega_k\,
    \mathcal{R}(\omega_k,\bm{\hat{k}};\omega_{k'},\bm{\hat k}')\,
	I(\omega_k,\bm{\hat{k}}) \nonumber \\
&\equiv&
	\left\{1 - \frac{1}{\rho(u)}\,\frac{\sum_j\rho(l_j)
    B_{l_ju}J(\omega_{ul_j})}{\sum_n\left[A_{ul_n} 
     + B_{ul_n}J(\omega_{ul_n})+ C_{ul_n} \right] + \Gamma_u}\right\}
	\varepsilon\apx{1}_0(\omega_{k'},\bm{\hat k}')
	+\varepsilon\apx{2}_0(\omega_{k'},\bm{\hat k}')\;.
\end{eqnarray}
\end{widetext}
The identification of the second addendum with the second-order
emissivity $\bm{\varepsilon}\apx{2}$ of \cite{Ca14} follows immediately
when we recognize that the total statistical width of the upper state
that enters the denominator ${\rm i}(\Omega_u-\Omega_{u'}^\ast)$ of
$\bm{\varepsilon}\apx{2}$ (in the presence of both inelastic and
elastic collisions; see Section~\ref{sec:coll}) is given by
\begin{equation}
	\epsilon_u=\textstyle\frac{1}{2}\sum_n
\left[A_{ul_n}+B_{ul_n}J(\omega_{ul_n})+C_{ul_n}\right]
	+\frac{1}{2}\Gamma_u\;.
\end{equation}
%
%
%

We now focus on the expression within curly braces in
equation~(\ref{eq:RHPRDepsilon}), and recall equation~(\ref{eq:rhou_one}) 
relating the populations of
the upper and lower states for the unpolarized multi-level atom of the 
$\Lambda$-type. We find
%
%
\begin{widetext}
\begin{eqnarray*} \label{eq:RHpar}
1&-&\frac{1}{\rho(u)}\,\frac{\sum_j\rho(l_j)
    B_{l_ju}J(\omega_{ul_j})}{\sum_n\left[A_{ul_n} 
     + B_{ul_n}J(\omega_{ul_n})+ C_{ul_n} \right] + \Gamma_u} \\
%
%
\mbox{\footnotesize\emph{using} (\ref{eq:gamma})+(\ref{eq:subst2})}\;\;
&=& 1 - \frac{1}{\rho(u)}\,(1-\gamma)\,
\frac{\sum_j\rho(l_j) B_{l_ju}J(\omega_{ul_j})}
	{\sum_n\left[A_{ul_n}+B_{ul_n}J(\omega_{ul_n})+C_{ul_n}\right]} \\
\mbox{\footnotesize\emph{using} (\ref{eq:rhou_one})}\;\;
&=& 1 - (1-\gamma)\sum_j 
	\frac{A_{ul_j}+B_{ul_j}J(\omega_{ul_j})+C_{ul_j}}
	{\sum_n\left[A_{ul_n}+B_{ul_n}J(\omega_{ul_n})+C_{ul_n}\right]}\,
\frac{B_{l_ju}J(\omega_{ul_j})}
{B_{l_ju}J(\omega_{ul_j})+C_{l_ju}} \\
%
\mbox{\footnotesize\emph{using} (\ref{eq:alpha_i})}\;\;
&=& 1 - (1 - \gamma)\sum_j\alpha_j
     \frac{B_{l_ju}J(\omega_{ul_j})}
     {B_{l_ju}J(\omega_{ul_j})+C_{l_ju}} \\
\mbox{\footnotesize\emph{using} (\ref{eq:w_c.mt})}\;\;
&=& 1 - (1 - \gamma)(1 - w_c) \\
&=& w_c + \gamma(1 - w_c)\;.
\end{eqnarray*}
\end{widetext}
We thus have demonstrated that equation~(\ref{eq:RHPRDepsilon}),
corresponding to the emissivity term of \cite{Ui01}, is equivalent to 
the intensity emissivity of our 
RT equation (\ref{eq:RT.coll}), when we adopt our definitions (\ref{eq:gamma}),
(\ref{eq:w_c.mt}), and (\ref{eq:alpha_i}).

\section{The general case of the polarized multi-term atom of the 
$\Lambda$-type}

The analysis in the previous section suggests a straightforward method 
for generalizing 
equation~(\ref{eq:RT.coll}) to include the effects of atomic polarization. 
In such case, one typically cannot write the separation between
$\rho(u)_{\rm rad}$ and $\rho(u)_{\rm coll}$ as simply as in
equation~(\ref{eq:rho_coll}), preventing a clear identification of the
partition coefficient $w_c$ (see the discussion immediately following 
the description of the diagrams in Figure~\ref{fig:diagrams}). 
On the other hand, it is known that in the limit of spectrally flat 
illumination, \emph{regardless of the presence of collisions} (whether 
inelastic and/or elastic), the total emissivity in the RT 
equation~(\ref{eq:RT.coll}) must be identical to $\bm{\varepsilon}\apx{1}$.

Relying on this fact, we introduce the quantity
\begin{equation} \label{eq:alt.eps1}
\bm{\tilde\varepsilon}\apx{1}\equiv\bm{\varepsilon}\apx{1}
	-\bm{\varepsilon}\apx{2}_{\rm f.s.}\;,
\end{equation}
where ``f.s.''\ stands for ''flat spectrum''. This is to indicate that the 
term $\bm{\varepsilon}\apx{2}_{\rm f.s.}$ is obtained from the general 
expression for $\bm{\varepsilon}\apx{2}$ by approximating the incoming
Stokes vector $\bm{S}(\omega_k,\bm{\hat{k}})$ with its spectral and angular
average, which can be expressed through the spherical tensors 
$J^K_Q(\omega_{ul})$ of the incident radiation field \citep{LL04}. Then 
the redistribution integral 
over the incident frequency in $\bm{\varepsilon}\apx{2}_{\rm f.s.}$ can be 
performed using equation~(15) of \cite{Ca14}.

It is important to observe that $\bm{\tilde\varepsilon}\apx{1}=0$ 
in the absence of collisions, owing to the fact that 
$\rho(u)\equiv\rho(u)_{\rm rad}$ in that case. Then, the equality 
of $\bm{\varepsilon}\apx{1}$ and of $\bm{\varepsilon}\apx{2}_{\rm f.s.}$ 
follows from the analysis of \cite{Ca14}. 
In the presence of collisions, instead, 
$\rho(u)\ne\rho(u)_{\rm rad}$ (and consequently, when 
elastic collisions are also present, 
$\rho(u)\ne(1-\gamma)\,\rho(u)_{\rm rad}$; 
cf.\ equation~(\ref{eq:subst})), and so 
$\bm{\tilde\varepsilon}\apx{1}\ne0$ in general.
Then, according to our diagrammatic analysis, $\bm{\tilde\varepsilon}\apx{1}$
accounts for the completely redistributed radiation produced via 
all possible excitation processes, with the exception of the radiative 
process leading to the coherent scattering of the incident 
radiation (Figure~\ref{fig:diagrams}.(ii)).\footnote{%
This is readily seen in the particular case of an unpolarized atom,
since
\begin{displaymath}
\rho(u)=\rho(u)_{\rm coll}+\rho(u)_{\rm rad}
=\rho(u)_{\rm coll}+\gamma\,\rho(u)_{\rm rad}
	+(1-\gamma)\,\rho(u)_{\rm rad}\;,
\end{displaymath}
and the last addendum corresponds exactly to the contribution of
$\bm{\varepsilon}\apx{2}_{\rm f.s.}$ (see discussion around
equation~(\ref{eq:subst})). Therefore the contribution of
$\bm{\tilde\varepsilon}\apx{1}$ corresponds to the sum 
$\rho(u)_{\rm coll}+\gamma\,\rho(u)_{\rm rad}$ (i.e., diagrams
(i) and (iv) of Figure~\ref{fig:diagrams}).}
It follows that $\bm{\tilde\varepsilon}\apx{1}$ automatically accounts 
also for the contribution of the upper-state atomic polarization to 
the completely redistributed radiation.

Therefore, rather than
attempting to define ad-hoc weights $w_c\apx{K}$ for the individual 
multipolar components $\rho^K_Q(u)$ of the atomic density matrix, 
in order to generalize equation~(\ref{eq:RT.coll}) to the case of
a polarized atom, we can simply operate the substitution
\begin{displaymath}
\bigl[	w_c + \gamma(1-w_c) \bigr]\,
		\bm{\varepsilon}\apx{1}\;
	\rightarrow\;\bm{\tilde\varepsilon}\apx{1}\;.
\end{displaymath}
With this generalization, the RT equation (\ref{eq:RT}) for the 
\emph{polarized multi-term atom of the $\Lambda$-type} can finally be 
given,
\begin{eqnarray} \label{eq:RT.gen}
&&\frac{d}{ds}\,\bm{S}(\omega_{k'},\bm{\hat k}') 
= -{\bf K}(\omega_{k'},\bm{\hat k}')\,
	\bm{S}(\omega_{k'},\bm{\hat k}')
	\nonumber \\
&&\kern .1in{}	+\Bigl[\bm{\varepsilon}\apx{1}(\omega_{k'},\bm{\hat k}')
	-\bm{\varepsilon}\apx{2}(\omega_{k'},\bm{\hat k}')_{\rm f.s.}
	\Bigr]
+\bm{\varepsilon}\apx{2}(\omega_{k'},\bm{\hat k}')\;,\kern .3in
\end{eqnarray}
where \emph{the first-order solution of the atomic density matrix should be
used to evaluate the emissivity and absorption coefficients.}
In particular, in the case of an unpolarized multi-level system, the
total contribution to the intensity emissivity in the above equation 
converges exactly to the emissivity 
$\epsilon_i(\omega_{k'},\bm{\hat k}')$ of 
\cite{Ui01} (cf.\ equation~(\ref{eq:RHPRDepsilon})), which we proved 
to be in agreement with 
our heuristic diagrammatic analysis of the branching between the 
CRD and PRD components of the scattered radiation (see 
Section~\ref{sec:coll}).

\section{Summary}

We discussed the problem of determining physically meaningful 
branching ratios between the contributions of completely and 
partially redistributed radiation (respectively, CRD and PRD) 
to the formation of polarized spectral lines in collisional plasmas.

Our analysis was based on the results of a recent diagrammatic theory 
of partially coherent scattering by polarized multi-term atoms of the
$\Lambda$-type \citep{Ca14,CM16}, which considers all radiation 
processes to second order of perturbation in the collisionless limit. 
The present work provides a heuristic extension of that theory to 
the collisional case, relying on an intuitive diagrammatic 
description of the interplay of radiation and collisional processes in 
spectral line formation to the same order of perturbation.

This work was motivated by the need to devise physically consistent 
numerical schemes for the modeling of scattering polarization
of spectral lines in realistic models of optically thick stellar 
atmospheres. These schemes are traditionally based on the iteration 
between the (local) solution of the statistical equilibrium (SE) of 
the plasma atoms with the (non-local) feedback of the emitted 
radiation on the plasma excitation.
The work of \cite{Ca14} provides the radiative transfer (RT) equation 
for partially coherent scattering, but the corresponding set of SE 
equations has not yet been derived. Hence the solution of the SE 
problem to perturbative first order \citep{LL04} must be used instead 
in practical cases \cite[e.g.,][]{CM16}. For this very reason, one must
introduce branching ratios between the CRD and PRD contributions to the 
RT equation of \cite{Ca14}, when redistribution effects in the modeled
problem are important, such as in the case of collisional plasmas.

In this paper we showed how this can be done self-consistently, without
impacting energy conservation, which is critical for the stability and
convergence of numerical schemes for polarized RT in optically thick
atmospheres. The guiding principle throughout is the fact that the 
second-order emissivity term $\bm{\varepsilon}\apx{2}$, which describes the 
coherent (in the broader sense of ``memory preserving'') scattering 
of radiation, must converge to the usual term of spontaneous emission in
the limit of spectrally structureless illumination \emph{and} in the
absence of collisions.

For the sake of simplicity, we formulated and solved first the problem
of the unpolarized two-level atom, to finally arrive at the 
generalization of the results to the case of the polarized multi-term 
atom of the $\Lambda$-type. For the intermediate case of the unpolarized 
multi-level atom, our results are found to agree with previous works 
on the subject \cite[e.g.][]{Ui01}.

One notable result of this study is that the expression of the 
$\bm{\varepsilon}\apx{2}$ emissivity term already accounts for the proper 
branching among the various radiative and collisional processes that 
determine the PRD contribution to the scattered radiation. Hence,
this emissivity term always enters the RT equation without the need for
a multiplicative weighting factor (cf.\ equation~(\ref{eq:RT.gen})). 
Only the CRD contribution to the RT equation must be modified in order 
not to violate energy conservation (cf.\ equation~(\ref{eq:alt.eps1})).

The final expression (\ref{eq:RT.gen}) of the RT equation for polarized 
scattering in a multi-term atom of the $\Lambda$-type was recently used 
to model the formation of the \ion{Mg}{2} h--k doublet in a magnetized 
atmosphere, taking into account both inelastic and elastic collisions, 
but neglecting stimulated emission \citep{dPA16}.

\begin{acknowledgments}
We thank R.\ Centeno (HAO) for internally reviewing the manuscript, and
for helpful comments that have improved the presentation of the material.
\end{acknowledgments}

\end{document}